\documentclass[twocolumn]{aastex62}
\usepackage{graphics} 
\usepackage{epsfig} 
\usepackage{times} 
\usepackage{amsmath} 
\usepackage{amssymb}  
\usepackage{xcolor}
\usepackage{subfigure}
\usepackage{bm}
\usepackage{amsfonts}
\usepackage{color}
\graphicspath{{./}}

\submitjournal{ApJ}

\shorttitle{Relativistic Particle Transport and Acceleration in Structured Plasma Turbulence}
\shortauthors{Pezzi et al.}

\begin{document}

\title{Relativistic Particle Transport and Acceleration in Structured Plasma Turbulence}

\correspondingauthor{Oreste Pezzi}
\email{oreste.pezzi@istp.cnr.it}

\author[0000-0002-7638-1706]{Oreste Pezzi}
\affiliation{Istituto per la Scienza e Tecnologia dei Plasmi, Consiglio Nazionale delle Ricerche, Via Amendola 122/D, I-70126 Bari, Italy}
\affiliation{Gran Sasso Science Institute, Viale F. Crispi 7, I-67100 L'Aquila, Italy}
\affiliation{INFN/Laboratori Nazionali del Gran Sasso, Via G. Acitelli 22, Assergi (AQ), Italy}

\author[0000-0003-2480-599X]{Pasquale Blasi}
\affiliation{Gran Sasso Science Institute, Viale F. Crispi 7, I-67100 L'Aquila, Italy}
\affiliation{INFN/Laboratori Nazionali del Gran Sasso, Via G. Acitelli 22, Assergi (AQ), Italy}

\author[0000-0001-7224-6024]{William H. Matthaeus}
\affiliation{Department of Physics and Astronomy, University of Delaware, Newark, Deleware 19716, USA}



\begin{abstract}
We discuss the phenomenon of energization of relativistic charged particles in three-dimensional (3D) incompressible MHD turbulence and the diffusive properties of the motion of the same particles. We show that the random electric field induced by turbulent plasma motion leads test particles moving in a simulated box to be accelerated in a stochastic way, a second order Fermi process. A small fraction of these particles happen to be trapped in large-scale structures, most likely formed due to the interaction of islands in the turbulence. Such particles get accelerated exponentially, provided their pitch angle satisfies some conditions. We discuss at length the characterization of the accelerating structure and the physical processes responsible for rapid acceleration. We also comment on the applicability of the results to realistic astrophysical turbulence. 
\end{abstract}



\section{Introduction} \label{sec:intro}

The fact that the cosmic-ray (CR) spectrum extends up to extremely high energies, as well as the difficulties encountered by standard acceleration mechanisms to energize particles up to such energies, has led to a wide search for ways to boost the maximum energy and/or alternative acceleration mechanisms. While we can safely say that CR acceleration at the shocks associated with supernova (SN) explosions is now confirmed experimentally, it is also true that convincing observational proof that SN remnants (SNRs) can accelerate protons to energies in excess of $\sim 100$ TeV is, thus far, missing. Even from the theoretical point of view, SNRs associated with Type Ia and ordinary core-collapse SNe may be connected to CR acceleration up to $\sim 100$ TeV, while higher energies require much more extreme conditions, perhaps to be found in rare and very energetic SN explosions \cite[]{cristofari2020}. 

While standard acceleration processes, such as the second-order Fermi process \cite[]{Fermi49} and diffusive shock acceleration \cite[]{1977ICRC11.132A,1977DoSSR2341306K,1978ApJ221L29B}, have received a lot of attention throughout the years, more recently, the acceleration of charged particles in realistic MHD turbulence has been attracting an increasing level of attention \citep[e.g.,][and references therein]{LazarianEA20}, with special emphasis on magnetic reconnection. Early on, attention focused on contributions to energization by MHD activity in special regions within a dynamic plasma \citep{Giovanelli47,dungey1953lxxvi}. Examination of diverse scenarios for energization has over time become increasingly varied and complex \citep{AmbrosianoEA88, DmitrukEA04-tp, DrakeEA06,DrakeEA09,KowalEA12-ptcles}, suggesting that a more general perspective may be led by a simpler view of the physical principles at work. 

In 2D configurations, there exist simple conservation laws that define and constrain particle orbits \citep{sonnerup1971adiabatic,MattEA84} so that when turbulent fluctuations are present, charged test particles can be confined and accelerated \citep{AmbrosianoEA88,DrakeEA06} in secondary magnetic flux structures or ``islands''. Initially applied to smaller-gyroradius particles such as electrons in the context of magnetic reconnection geometries,
this idea was further developed by \citet{OkaEA10}, who noted that coalescence in a multiple-island context can efficiently accelerate electrons due to ``anti-reconnection'' electric fields associated with such mergers.  Furthermore, when applied to ``pickup'' of protons in reconnection jets \citep{DrakeEA09} that subsequently feed into a Fermi process as multiple magnetic islands contract and merge, a more complex proton energization process may be developed involving multiple islands \citep{DrakeEA10}. More recently, \citet{trotta2020fast} showed that transrelativistic electrons can be significantly accelerated while trapped in turbulent structures and experiencing curvature drift. Such a behavior, obtained in a 2D configuration where the background turbulent plasma is modeled with a hybrid particle-in-cell method while energetic electrons are treated as test particles, has also been verified at different intensities of the turbulent fluctuations. The basic physical elements of these models were formalized in transport theories \citep{Zank2014transport,leRouxEA15,leroux2018self,leroux2019modeling} that facilitate applications (see also the recent reviews by \citet{khabarova2021currentsheets, pezzi2021currentsheets}). 

Complementary to these developments, there has been a parallel line of studies that begin with the premise that charged particle energization might be treated as being due to plasma dynamics that from the onset is complex and, in fact, turbulent. Early efforts demonstrated the feasibility of efficient turbulent acceleration \citep{DmitrukEA03-testpart} in three dimensions, although numerical limitations such as small system size and lack of resonant power at numerical grid scales (however see \citep{LeheEA09}) cast doubt on conclusions concerning scaling laws. 

Numerical experiments have improved ever since \citep{DmitrukEA04-tp,KowalEA12-ptcles,DalenaEA14}, including turbulence effects associated with current sheets and reconnection, contracting islands, and proliferation of plasmoids and/or secondary islands \citep{KowalEA11,KowalEA12-ptcles}. Similar effects have been characterized in relativistic plasmas \citep{hoshino2012stochastic, zhdankin2017kinetic,zhdankin2019electron,ComissoSironi18,GuoEA20,KilianEA20}. 

Most important, it has been understood that the initial efforts to characterize acceleration in or around reconnection regions focused on the physical processes responsible for extracting particles from the thermal background and injecting them into some energization process at work on larger scales. Yet, the particle velocity, the thermal velocity, and the Alfv\'en speed were of the same order of magnitude, making it very difficult to identify the nature of the acceleration mechanism (first or second order) and the scaling laws that could be used for higher-energy particles. 

Injecting test particles in a 2D or 3D simulation box of MHD turbulence has been an independent method to investigate both the transport of such particles and their energization, although in some of the previous articles (e.g. \citep{KowalEA11,KowalEA12-ptcles}) the Larmor radius of the particles at the beginning of the simulation was chosen to be much smaller than the grid spatial spacing, which makes transport unrealistic, due to the lack of resonance with the turbulence and thereby unreliable diffusive transport. 

When self-consistent broadband turbulence is involved, particle energization becomes complex due to interactions with the internal structure of large- and small-scale flux tubes, as well as the current sheets, vortices, and reconnection sites that typify flux-tube boundaries and their mutual interactions. Common to a number of these treatments of energization including turbulence is the role of direct acceleration for particles of smaller Larmor radii, transitioning to the involvement of perpendicular acceleration at larger energy \citep{DmitrukEA04-tp,DalenaEA14,ComissoSironi18,comisso2021pitch,trotta2020fast}.

Although some works indicated that high energies can be also attained by second-order processes (e.g., \citet{ArznerEA06,SioulasEA20}), the role of temporary trapping has been highlighted in various contexts as it can dramatically influence both transport and acceleration \citep{KowalEA11,KowalEA12-ptcles,TooprakaiEA16,DalenaEA14}. Indeed, the ubiquity of turbulent coherent structures and islands/plasmoids/flux ropes produced by magnetic reconnection makes the potential of the trapping mechanism significant for different systems. In the solar and stellar coronae, magnetic loops may provide sufficient conditions to entrap particles \citep{vlahos2018particle}. In the interplanetary medium, there are observations supporting the idea that particles are locally accelerated when trapped in merging or coalescing islands \citep{khabarova2017energetic, MalandrakiEA19}. Particle trapping may also provide a source for particle reacceleration downstream of shocks \citep{zank2015diffusive,nakanotani2021interaction}. Other systems, e.g. the intracluster medium, could also encompass such a mechanism although observations are usually explained in terms of second-order processes \citep{vazza2017turbulence, brunetti2020second}. The importance of trapping for accelerating particles has been also recently highlighted by \citet{lemoine2021particle} in the context of strong and intermittent turbulence by relating the energization process and the gradients of the bulk velocity and magnetic fields.

Intuitively, trapping can enhance acceleration when appropriate electric fields are encountered, but it can also inhibit stochastic acceleration when the required transport is thwarted. The full range of possibilities for such effects remains to be exhaustively explored. The study of \citet[][see also \citet{KowalEA12-ptcles}]{KowalEA11} is instructive on several salient features. In these numerical experiments, the authors observe two small magnetic islands merging with a central elongated island. A selected particle gains considerable energy after entering the system by following field lines through one small island, eventually becoming trapped within the large island, circuiting it numerous times, and gaining energy exponentially and mainly when passing near the region of merger with the small island. 

These conclusions were illustrated in detail only for 2D turbulence in a configuration that was optimized to create reconnecting islands. However, the test particles injected in a snapshot of the simulation were initially subgrid, which means that their transport could not be described in a realistic way. Eventually, the energy of the particles, after an exponential increase, reaches the regime in which resonances could in principle be relevant for particle transport, but this phenomenon was not discussed. Despite these shortcomings, 
this approach demonstrates clearly the complex interplay between the transport effects that entrain the particle within the island and near the acceleration region, along with the special circumstances that support the electric field responsible for the energization itself.

In the present paper, we advance such a scenario for strong turbulence and particle energization in the highly relativistic particle regime, having in mind the implications that the process may have for the acceleration of very-high-energy CRs. In particular, we continue the examination of these complex interactions of charged particles with turbulence by performing relativistic test-particle simulations with a turbulent electromagnetic field produced by means of 3D MHD simulations, not specifically devised to produce reconnection regions. Our focus is on clarifying of the nature of the trapping or entrainment that leads particles to rapid acceleration to the highest energies.

We find that the bulk of the test particles injected in the simulation box went through a secular-second order acceleration due to the random plasma motions, which in turn induced random electric fields. The interaction of the particles with these plasma motions leads to stochastic energy change, which we characterize in terms of plasma properties and manage to associate with a diffusive motion in momentum space. The transport of the same particles in physical space is also found to be well described through a diffusive motion. The latter is in a range of scales where the anisotropic cascade of the turbulence with respect to the local magnetic field does not seem to have a visible effect as yet. 

In addition to this acceleration process that is clearly at the second order in the quantity $v_A/c\ll 1$, we also identify a small fraction of particles that manage to get trapped in selected regions associated with the interaction between flux tubes. These particles all have pitch-angle cosine very close to zero, a necessary condition for trapping, and go through an exponential phase of energy increase. We provide a characterization of these regions in terms of physical observables that could be measured in the simulation. We also build a simple model of the region where this phenomenon occurs and manage to reproduce the main properties of the acceleration process and the time scales involved. The acceleration process is similar to a first-order mechanism in which the particle trajectory in the plane perpendicular to the local magnetic field encounters a gradient of plasma velocity (i.e. of the induced electric field). 

The paper is structured as follows. In Sect. \ref{sec:background} we present the MHD simulations adopted in the present work, while in Sect. \ref{sec:testparticle} we describe the test-particle code and the first results obtained in terms of physical space transport. Then, in Sect. \ref{sec:results} we focus on the main numerical outcomes of the work concerning particle energization. Sect. \ref{sec:discussion} discusses a simple model of the acceleration region and derives the main properties of the acceleration mechanism and the main time scales involved. Moreover, we discuss the implication of our findings for astrophysical systems. Finally, in Sect. \ref{sec:conclusions} we conclude by summarizing our results and illustrating future developments.

\section{MHD Simulation Background} \label{sec:background}

In order to study the transport and acceleration of charged test particles, we follow particle evolution 
in electromagnetic fields obtained through incompressible three-dimensional MHD simulations. These simulations solve the following set of equations:
\begin{eqnarray}
\frac{\partial {\bm u}}{\partial t} + \left({\bm u} \cdot \nabla\right) {\bm u} &&= - \frac{1}{\rho}\nabla P + \frac{1}{\rho}{\bm j} \times {\bm B} + \nu \nabla^2 {\bm u} \label{eq:MHD1}\\ 
\frac{\partial {\bm B}}{\partial t} + \left({\bm u} \cdot \nabla\right) {\bm B} &&= \left({\bm B} \cdot \nabla\right) {\bm u} + \eta \nabla^2 {\bm B} \\ 
\nabla \cdot {\bm u} &&= \nabla \cdot {\bm B} = 0 \label{eq:MHD2}
\end{eqnarray}
where ${\bm u}({\bm r},t)$ is the magnetofluid speed composed only of its fluctuating part, and ${\bm B}({\bm r},t)$ is the magnetic field that is decomposed into a uniform mean $B_0$ and a zero-mean fluctuation $\bm b$, ${\bm B}({\bm r},t) = \bm{B_0} + \bm{b}({\bm r},t)=B_0 {\bm e}_z + \bm{b}({\bm r},t)$. Furthermore, $P$ is the thermal pressure, and $\rho$ is the magnetofluid density. The current density is ${\bm j}=\nabla \times {\bm B}$, while $\nu$ and $\eta$ are the viscosity and resistivity, respectively. The flow is incompressible $\nabla \cdot {\bm u} =0$, and 
the density is uniform $\rho={\rm const}$. 

Lengths, time, and velocities in Equations (\ref{eq:MHD1}--\ref{eq:MHD2}) are respectively normalized to a typical length $L_A$, time $t_A$ and to the Alfv\'en speed $v_A=L_A/t_A=\bar{B}/\sqrt{4\pi m_p\bar{n}}$, where $\bar{B}$ and $\bar{n}$ are reference values for the magnetic field and for the background number density. We here adopt $L_A=81.5\,  {\rm pc}$, corresponding to $L_{\rm box}=512 \, {\rm pc}$, while ${\bar B}=1\, \mu G$ and ${\bar n}=1\, {\rm cm}^{-3}$. Unless specified, hereafter we assume normalized variables. 

\begin{figure}[htb!]
  \centering
  \includegraphics[width=\columnwidth]{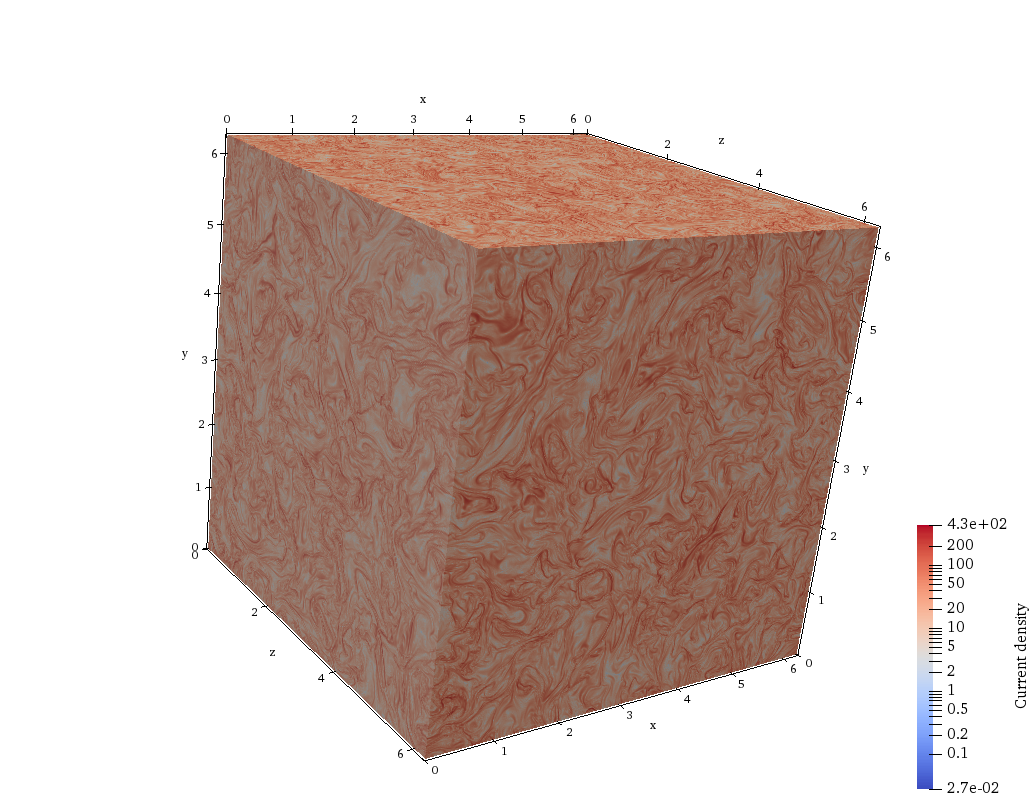}
  \caption{Rendering of the 
  current density $j^2({\bm r})$  shows a plethora of intermittent coherent structures. Such structures form a template for the possibility of rare acceleration events.}
  \label{fig:j2B0=0}
\end{figure}

Equations (\ref{eq:MHD1}--\ref{eq:MHD2}) are solved in a 3D Cartesian periodic box of size $L_{\rm box}=2\pi$, with spatial 
resolution $N_x=N_y=N_z=1024$ adopting a pseudo-spectral method in a Fourier basis. The time advancement is performed with a second-order Runge-Kutta scheme and the 2/3 rule for spatial dealiasing is chosen \citep{PattersonOrszag71}.
Small values of resistivity and viscosity $\eta=\nu=2\times 10^{-4}$
are introduced to define the well-resolved spectral domain. The dissipative wavenumber $k_{\rm diss}$ 
(the reciprocal of the Kolmogorov length scale) for the considered runs is always smaller by a factor 2 than the maximum resolved wavenumber $k_{\rm max}$ \citep[for further details, see][]{BandyopadhyayEA18-prx}.

\begin{figure}[htb!]
  \centering
  \includegraphics[width=\columnwidth]{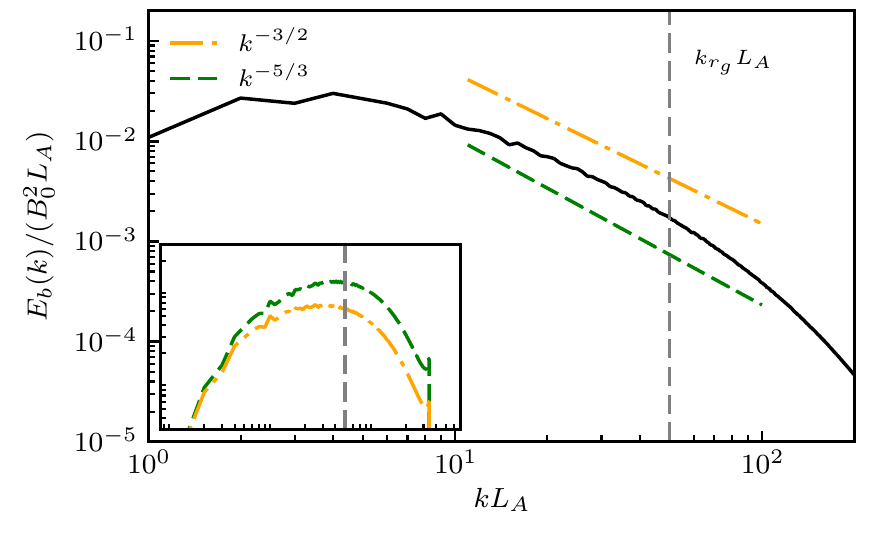}
  \caption{Omnidirectional spectrum of the magnetic energy. The green dashed (orange dotted-dashed) line shows the Kolmogorov (Kraichnan) prediction. The small inset displays the magnetic energy spectrum compensated by the Kolmogorov (green dashed) and the Kraichnan (orange dotted-dashed) slope. The gray dashed vertical lines indicate the wavenumber associated with the initial particle gyroradius. }
  \label{fig:MagneticSpectrum}
\end{figure}

Large-scale uncorrelated fluctuations of ${\bm u}$ and ${\bm b}$ are introduced at $t=0$ and turbulence develops, producing small-scale fluctuations. We focused here on the case with $u_{\rm rms}=b_{\rm rms}=1$ and $B_0=0$. The role of a finite background magnetic field and compressibility will be discussed in a separate forthcoming work.
We then selected the time instant at which the turbulent activity is strongest (i.e. highest dissipation). The complex and highly structured pattern of the turbulence is displayed in Figure \ref{fig:j2B0=0}, showing the contour plot of ${\bm j}$ in the 3D domain. Vortices and magnetic islands, as well as intense current sheets where magnetic reconnection may be at work, naturally emerge as elementary structures of the turbulent flow. The omnidirectional spectrum of magnetic energy (Figure \ref{fig:MagneticSpectrum}) indicates that an inertial range, whose length is about a decade in wavenumber space, develops before dissipative effects steepen the spectrum at higher wavenumber $k$.
In the inertial range, the slope is rather compatible with either the
Kolmogorov or Kraichnan predictions; 
these are respectively displayed in green and orange dashed lines in Figure \ref{fig:MagneticSpectrum} (see also the inset in the same figure). By numerical evaluating the correlation length $l_c$ of the magnetic field, we find $l_c=0.218$ ($l_c=17.7\, {\rm pc}$ in physical units), corresponding to protons with energy $E\sim 16\, {\rm PeV}$ in the typical field ${\bar B}$.

\section{Methods: test-particle propagation details}\label{sec:testparticle}
We numerically integrate the motion equations of $N_p=10^{5}$ relativistic test particles of positive charge $e$ and mass $m_p$ moving in the turbulent electromagnetic field obtained by means of the incompressible MHD simulations described above. The normalized particle equations of motion are
\begin{eqnarray}
&&\frac{d \bm{x}}{dt} = {\bm v} \label{eq:rsc}\\
&&\frac{d {\bm p}}{dt} = \alpha \left( {\bm E} + {\bm v} \times {\bm B}   \right) \label{eq:usc}
\end{eqnarray}
where ${\bm x}=(x,y,z)$, ${\bm v}$, and ${\bm p}=\gamma {\bm v}$ are the particle position, velocity, and momentum, while ${\bm E}$ and ${\bm B}$ are the electric and magnetic fields. Equations (\ref{eq:rsc}-\ref{eq:usc}) are scaled analogously to MHD simulations. In normalized units, the Lorentz factor reads $\gamma = 1/\sqrt{1 - (\beta_A v)^2} = \sqrt{1 + (\beta_A p)^2}$, where 
$\beta_A = v_A/c$. The electric field in Equation (\ref{eq:usc}) is derived through Ohm's law: ${\bm E}= -{\bm u}\times{\bm B}+\eta {\bm j}$.

The parameter $\alpha = t_A \Omega_{0}$, where $\Omega_0=e \bar{B}/m_p c$ is the proton cyclotron frequency, can be easily rewritten as $\alpha=L_A/d_p$, with $d_p$ the proton skin depth of the background plasma. $\alpha$ is thus connected to the extension of the inertial range of the turbulence with respect to kinetic, dissipative scales \citep{DmitrukEA04-tp,GonzalezEA16}. In a $\beta_p\sim 1$ plasma (with $\beta_p$ the thermal to magnetic pressure ratio), the parameter $\alpha$ corresponds to the inverse of the normalized gyroradius of nonrelativistic  
particles moving with speed $\sim v_A$. Previous works considering the injection of thermal particles into the acceleration region were hence forced to reduce $\alpha$ to much smaller and computationally feasible values.
Such a requirement provides particles with a gyroradius at least larger than the grid size, so that resonant scattering might be properly taken into account. On the other hand, relativistic particles moving at the speed of light have a much larger gyroradius because $\gamma\gg 1$, thus removing the constraint on the value of $\alpha$. For the 
parameters described above, $\alpha\sim10^{12}$ and $\beta_A\sim10^{-5}$.
To save computational resources, we only artificially increase $\beta_A=5\times 10^{-2}$.

Because we are interested in the energization of relativistic particles moving in a nonrelativistic environment, we assume stationary electromagnetic fields, i.e. $\partial {\bm B}/\partial t = \partial {\bm E}/\partial t =0$ (magnetostatic approximation), and we consider a static snapshot of these fields when turbulence is fully developed.

Eqs. (\ref{eq:rsc}--\ref{eq:usc}) are integrated by adopting the relativistic Boris method \citep{ripperda2018comprehensive,dundovic2020novel}. The electric and magnetic fields are interpolated at the particle position through a trilinear interpolation method \citep{birdsall2004plasma}. We verified that the results presented here are not affected by adopting a more accurate yet significantly slower 3D cubic spline method (not shown here). Particles are injected homogeneously throughout the computational box at a given energy and with isotropic velocity direction on the unit 3D sphere. The time step is set to $1/50$ of the initial gyroperiod.

Most of the results here adopt the initial gyroradius to be $r_{g,0}\simeq 0.1 l_c=0.02$, corresponding to $E_0\simeq 1.6\, {\rm PeV}$. The resonant wavenumber $k_{r_g}=1/r_g = 50$, reported in Figure \ref{fig:MagneticSpectrum} with a vertical dashed gray line, resides in the inertial range of turbulence, and it is also quite far from the dissipative scales where the resistive electric field is expected to become important. This ensures that the acceleration process studied here is mainly driven by the inductive electric field, this being the most relevant term for analyzing the energization of relativistic particles whose gyroradius is much larger than the typical length where dissipative and resistive effects are expected to steepen the magnetic spectrum. To double-check, we also verified
that our results are not affected by the resistive field, in that if we exclude the resistive component from the computation of the electric field, the energization process is basically unchanged. This shows that for the high-energy particles we are interested in, namely when the Larmor radius exceeds the thickness of the reconnection regions, the energization is not due to the resistive fields but rather to the induced electric fields due to the plasma motion. 

Although here we are most interested in how particles react to electric fields in the simulation box, it is first worth studying how particles move in the magnetic field, especially to confirm that we find diffusive motion and to identify possible differences with respect to cases where turbulence is synthetic rather being the result of an MHD simulation. 

In order to study particle transport in physical space, we performed a subset of test-particle simulations by excluding the electric field. A diffusive regime after a ballistic transient is always recovered. When reaching the diffusive plateau, the isotropic diffusion coefficient is computed as $D_{\rm iso}=(D_{xx}+D_{yy}+D_{zz})/3$ with
 \begin{equation}
     D_{xx}(\Delta t) = \frac{\langle (\Delta x(\Delta t))^2 \rangle}{2 \Delta t}\, .
     \label{eq:D_run}
 \end{equation}
 
\begin{figure}[htb!]
  \centering
  \includegraphics[width=\columnwidth]{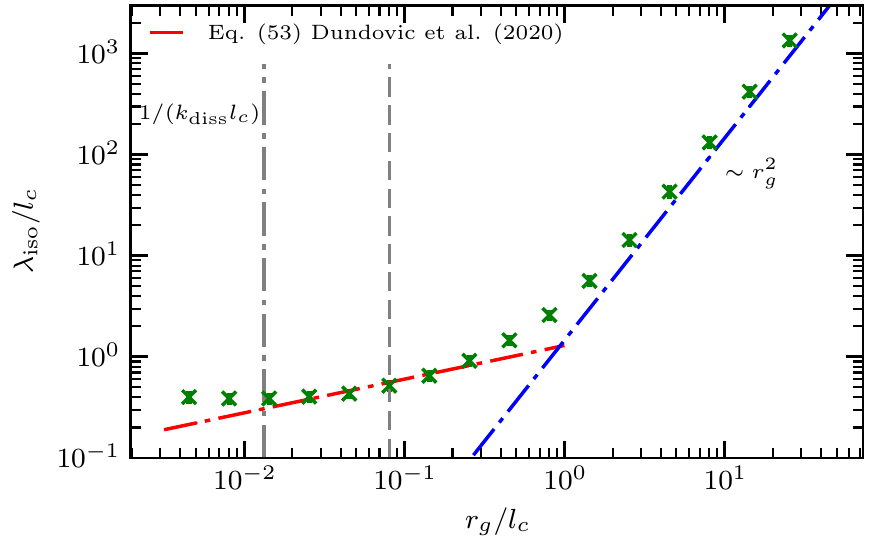}
  \caption{Mean free path $\lambda_{\rm iso}$ as a function of the particle gyroradius. The vertical dashed and dotted-dashed gray lines correspond to the particle gyroradius adopted here and to the gyroradius corresponding to the dissipative scale, respectively. The blue dotted-dashed line refers to the QLT prediction $D_{\rm iso} \sim r_g^2$ obtained for $r_g\gtrsim l_c$. The red dashed line reports the prediction of \citet{SubediEA17}  (Equation (53) of \citet{dundovic2020novel}).}
  \label{fig:Diso-sat}
\end{figure}

Figure \ref{fig:Diso-sat} shows the isotropic mean free path $\lambda_{\rm iso}=3 D_{\rm iso}/c$ as a function of the particle gyroradius $r_g$, normalized in the usual way to the correlation scale $l_c$. The typical behavior of the path length as a function of energy is the same as that found in synthetic turbulence, with a low-energy trend that reflects the shape expected from a given isotropic power spectrum \citep{SubediEA17,dundovic2020novel}. In particular, the dotted-dashed red line in Figure
\ref{fig:Diso-sat} implements Equation (53) of \citet{dundovic2020novel} for the Kolmogorov case, where $l_{\rm iso}$ is the bend-over scale of the synthetic model in  \citet{dundovic2020novel}, being $l_{\rm iso}\sim 2 l_c$. At variance with synthetic models of the turbulent field, the slope of the isotropic power spectrum is not very well defined here because of the limited dynamical range (see Figure \ref{fig:MagneticSpectrum}). 

At high energies, when the gyroradius satisfies the condition $r_g>l_c$, the diffusion coefficient becomes weakly dependent upon the power spectrum, $D_{\rm iso}\sim r_g^2$ (blue dotted-dashed line). For gyroradii $r_g/l_c<0.01$, several numerical effects, especially dissipation, limit the validity of the approach by reducing the power available in the form of modes that the particle gyration can resonate with. As a result, the numerically computed path length departs from the dotted-dashed red line at such low energies. The vertical dashed gray line identifies the energy of the particles used below for our investigation of energization. 

This correspondence to the isotropic spatial diffusion theory is itself a result of some significance: on one hand, it validates the approach of \citet{SubediEA17} and \citet{dundovic2020novel}
with a ``realistic'' turbulent and intermittent magnetic field obtained from the numerical evolution of MHD equations. On the other hand, this may be considered somewhat surprising because the anisotropic cascade development with respect to the local magnetic field, expected in MHD \citep{GS1994}, seems to have little effect on the diffusion properties. It is likely that the effects of the anisotropic cascade are not fully developed as yet, due to the limited dynamical range imposed by the numerical constraints. These results seem to be in good agreement with those of \cite{Cohet2016}.

\section{Results on particle energization}\label{sec:results}

The turbulent motion of the plasma in the simulation box leads to the unavoidable creation of inductive electric fields, which are expected to have random orientations. As such, their presence is expected to lead to changes in the energy of the particles, due to the presence of such inductive electric fields in the Lorentz force. This effect is expected to cause the energy of a particle to increase or decrease depending on the relative orientation of the particle momentum and the local electric field, namely a typical second-order phenomenon. On the other hand, the complex structure of MHD turbulence is known to trigger additional phenomena that may lead to more rapid energization of the particles  (see for instance \citet{KowalEA11}). Here we investigate all these phenomena in great detail, stressing that the simulation was not carried out to maximize the formation of reconnection regions or other peculiar structures. The phenomena we see are, in this sense, very generic. 

We find that in, addition to an overall second-order stochastic acceleration mechanism, a first-order process is at work as well, due to the temporary trapping of particles in coherent structures. 

\subsection{Stochastic energization}
\begin{figure}[!thb]
  \centering
  \includegraphics[width=\columnwidth]{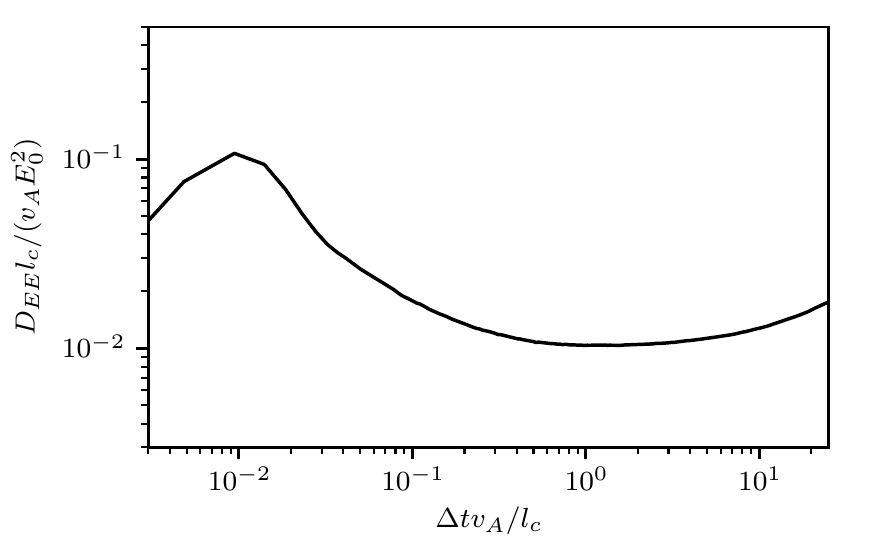}
  \caption{Running energy diffusion coefficient $D_{EE}$ as a function of $\Delta t$.}
  \label{fig:DE-run}
\end{figure}

Figure \ref{fig:DE-run} displays the running energy diffusion coefficient as a function of energy, assuming that in fact the motion of the particles can be described in terms of a random walk in momentum space:
\begin{equation}
    D_{EE} = \frac{\langle (\Delta E(\Delta t))^2\rangle}{2 \Delta t}.
    \label{eq:DEE}
\end{equation}
It is evident that after a transient, the energy diffusion coefficient saturates at a roughly constant value. This implies the presence of diffusion in energy space, thus revealing the typical nature of a second-order process \citep{Ostrowski97}. The energy diffusion coefficient $D_{EE}\simeq 0.01 v_A E_0^2/l_c$ implies a characteristic time for the energy diffusion process $\tau_{\rm diff,E}=E^2/D_{EE}\sim 10^2 l_c/v_A$ that is consistent with the large time scale for growth of the average energy that occurs at late times in our simulations (not shown here). The slight increase recovered in $D_{EE}$ for very large $\Delta t$ may be due to the fact that the average gyroradius starts to increase on this timescale due to other processes (see below), thus making $D_{EE}$ move away from the plateau.

\begin{figure}[thb!]
  \centering
  \includegraphics[width=\columnwidth]{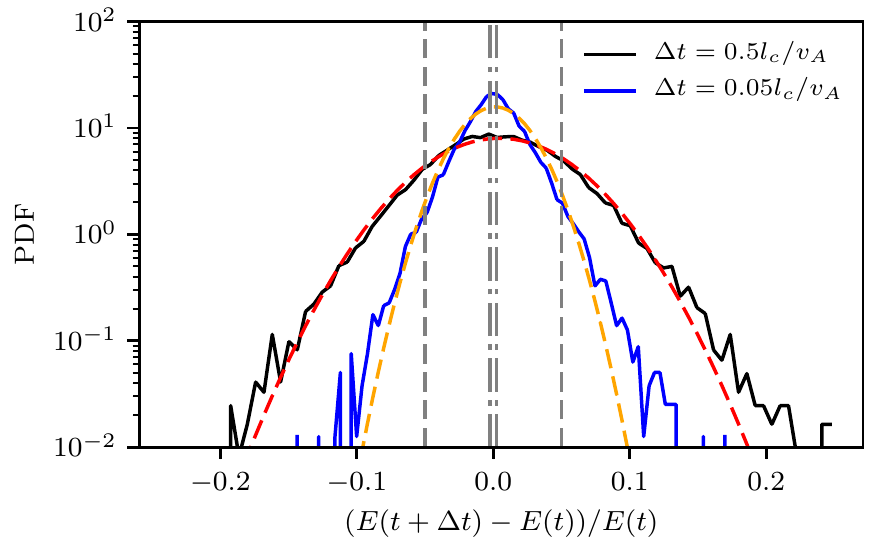}
\caption{Probability density functions (PDFs) of $\Delta E/E$ with $\Delta t=0.05 l_c/v_A$ (blue) and $\Delta t=0.5 l_c/v_A$ (black). The red and orange dashed curves correspond to the associated Gaussian distributions. Dotted-dashed and dashed gray lines indicate the values of $\beta_A^2$ and $\beta_A$, respectively. }
  \label{fig:dEtoE-dist}
\end{figure}

Another signature of an active stochastic energization mechanism comes by looking at the probability density functions (PDFs) of the relative energy gain $\Delta E(\Delta t)/E$, displayed in Figure \ref{fig:dEtoE-dist} for $\Delta t= 0.05 l_c/v_A$ (blue) and $\Delta t=0.5 l_c/v_A$ (black). The PDFs have been computed averaging on the initial time instants $t$ up to $t=t_{\rm max}\simeq 22 l_c/v_A$. Particles are likely to undergo both increases and decreases in energy. Although the distribution is peaked at small values, larger changes of energy up to $\gtrsim \beta_A$ (dashed gray lines in Figure \ref{fig:dEtoE-dist}) are allowed. The distribution functions are skewed toward the positive value of energy changes because the standardized skewness is $\tilde{s}=0.20$ for $\Delta t= 0.05 l_c/v_A$ and $\tilde{s}=0.12$ for $\Delta t= 0.5 l_c/v_A$, where $\tilde{s}=s/\sigma^3$. Here,
$\sigma$ and $s$ are, respectively, the standard deviation and the skewness (third-order moment) of the distribution function. The distribution function of the relative energy gains is manifestly non-Maxwellian for small $\Delta t$ and tends to recover the Maxwellian shape for larger $\Delta t$. Indeed, the kurtosis $\kappa$---defined as the fourth-order moment of the PDF normalized by $\sigma^4$--- is $\kappa = 4.84$ and $\kappa=3.69$ for $\Delta t= 0.05 l_c/v_A$ and $\Delta t= 0.5 l_c/v_A$, respectively.

The presence of positively skewed PDFs indicates that energy increase is more favorable than energy decrease. This provides the secular direction of the process, leading to a net energy gain. 

Clearly, all acceleration processes are at work simultaneously, and it is not trivial to discriminate among them by looking at a collection of particles: What we can say is that basically all test particles launched in the simulation box suffer from the second-order process illustrated above. As we discuss below, a small fraction of particles happen to be trapped in selected structures and get energized through a first-order process. It is not clear to what extent these few particles can affect the shape of the high-energy-gain tail of the PDF shown in Figure \ref{fig:dEtoE-dist}. As a consequence of the fact that only a few particles experience trapping, it is in general rather difficult to evaluate statistical properties of the population of trapped particles, such as the transport coefficients. In future work, we plan to explore the potential of novel methods, for instance, those commonly adopted in biophysics and based on single-particle trajectories \citep{golding2004RNA,saxton2012wanted,trotta2020particle}.

\subsection{Particle trapping in coherent structures: First-order acceleration}

\begin{figure}[hbt!]
  \centering
  \includegraphics[width=\columnwidth]{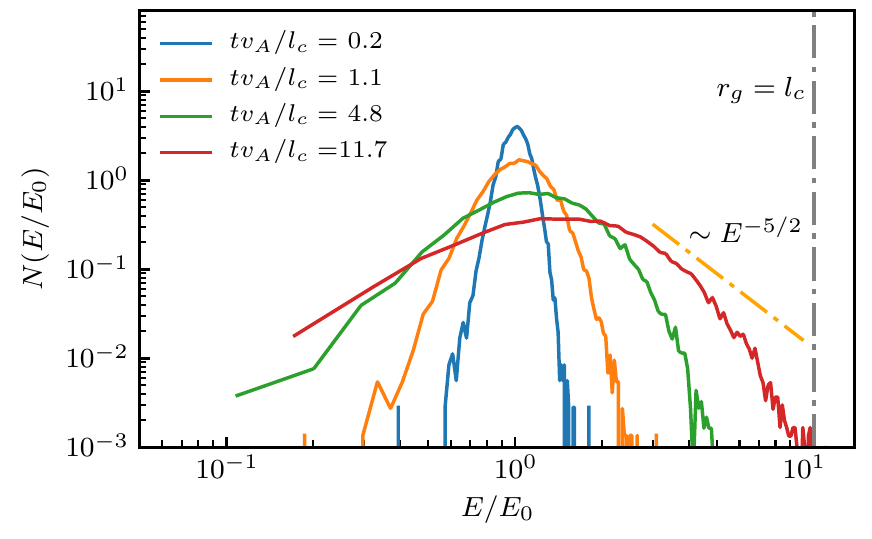}
  \caption{Particle PDFs at different time instants showing the energization process. The vertical dotted-dashed gray line highlights $r_g=l_c$, while the orange dotted-dashed line displays the ${-5/2}$ power-law slope.}
  \label{fig:ParticleEnergySpectra}
\end{figure}

In Figure \ref{fig:ParticleEnergySpectra} we show the spectra of particles in the simulation box, after a time $t$ indicated in the figure. A few comments are in order: (1) Particles are injected at energy $E_0$ which, for the natural units adopted here, corresponds to $E_0=1.6\, {\rm PeV}$. (2) As time evolves, the second-order process leads to a broadening of the distribution function, namely there are both particles losing energy and particles gaining energy. On average, however, the particle energy increases as one can see by noticing that the peak of the distribution moves toward higher energies. (3) Contemporaneously an approximate power law is created at high energies that eventually extends to particles with energies such that $r_g\simeq l_c$. This typically happens at times $t\gtrsim 10l_c/v_A$.

For the sake of comparison, we also report in Figure  \ref{fig:ParticleEnergySpectra} a line indicating the spectrum $\propto E^{-5/2}$, which was predicted and observed by different groups, although for very different systems and with different qualitative premises.

The pioneering numerical simulations of \citet{AmbrosianoEA88} found evidence for accelerated particles with a power-law tail with a slope compatible with $-5/2$ using two-dimensional simulations. As we discuss below, the dimensionality of the problem is very important in assessing the efficiency of trapping processes that are responsible for particle energization. Moreover, \citet{AmbrosianoEA88} focused on the extraction of particles from the thermal bath, for which the particle velocity remains close to that of the background thermal particles. As we discuss below, the acceleration, the trapping, and escape from the acceleration region work in somewhat different ways for relativistic particles. Moreover, as it is well known, the slope in energy should not be the same in any case for relativistic and nonrelativistic particles: This is so even for particles accelerated at a strong shock, for which the spectrum of accelerated particles in momentum is $f(p)\propto p^{-4}$, but when expressed in terms of energy, it is $N(E)\propto E^{-2}$ for relativistic particles and $\propto E^{-3/2}$ for nonrelativistic particles. 

The slope $-5/2$ was also predicted by \citet{degouveiadalpino2005production} and \citet{delvalle2016properties}, where a shock-like toy model was introduced to describe a reconnection region: The particles would be advected into the reconnection region with the inflowing plasma and would be expelled (the analog of escape to downstream in the case of a shock) at the speed of the reconnection exhaust. This determines a sort of universal spectrum for the accelerated particles. Such universality was later criticized by \citet{drury2012firstorder}. In fact, for particles that are being energized, it is unlikely that the velocity of the exhaust plays any role in the shaping of the spectrum of accelerated particles because the particles' Larmor radius becomes quickly larger than the thickness of the current sheet. As we mentioned above, the spectrum $E^{-5/2}$ is shown in Figure  \ref{fig:ParticleEnergySpectra} only as a reference, while it appears to be asymptotically reached in our simulations only for exceedingly long times compared with the dynamical time of the MHD turbulence.

\begin{figure}[htp!]
  \centering
  \includegraphics[width=\columnwidth]{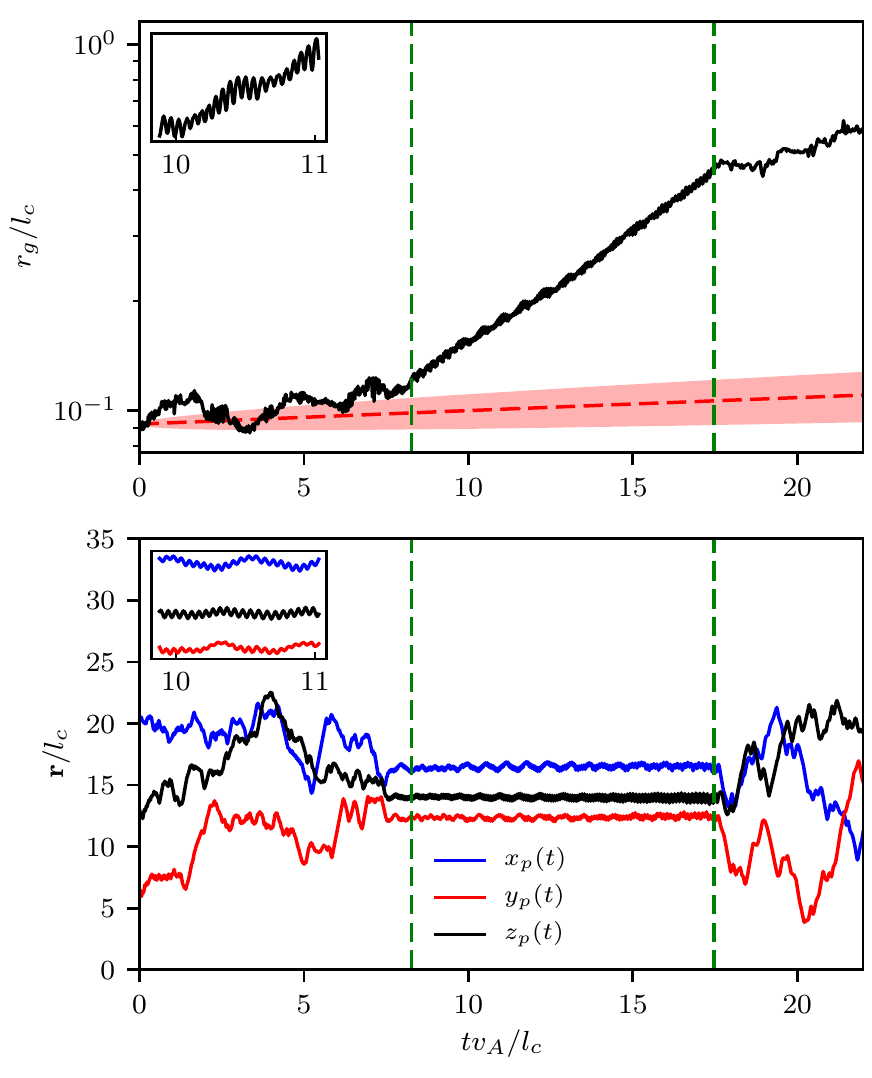}
  \caption{Typical behavior of a trapped particle showing exponential growth of energy that occurs within the two vertical green dashed lines. The particle gyroradius for the trapped particles increases exponentially over a time-scale $\tau \gtrsim 10 l_c/v_A$. This growth is much faster than the growth of the averaged gyroradius, where the average is performed on the full ensemble of test particles (red dashed line). The dashed red area represents the standard deviation of the averaged gyroradius. The bottom panel shows the particle trajectory illustrating that the particle is trapped. }
  \label{fig:Expgrowth}
\end{figure}

In the perspective of understanding the nature of the acceleration processes at work, in the top panel of Figure \ref{fig:Expgrowth}, we show the temporal evolution of the gyroradius averaged on the full particle ensemble (red dashed line), while the red shadowed area corresponds to the standard deviation of the averaged gyroradius. We clearly see that there is a secular increase in the particles' energy, which we attribute to the random interaction with the inductive electric fields in the simulation box. The black curve shows the temporal evolution of the particle gyroradius that, after a time $T\sim 22 l_c/v_A$, turns out to be the most energetic particle in the simulation. It is worth noticing that for early times, namely on the left side of the first vertical dashed line ($t\approx 8 l_c/v_A$), the fluctuations in the particle gyroradius (or its energy) are compatible with the fluctuations expected based on the bulk of the particles in the simulation (shaded area). Moreover, the particle energy is clearly carrying out a random walk, in that it increases and decreases, a typical feature of a second-order process. 

At a time $t\approx 8 l_c/v_A$ an exponential increase of the particle energy starts (the plot is in lin-log scale) and lasts for about $\sim 10 l_c/v_A$. The end of this period is marked by the rightmost vertical dashed line ($t\approx 17.5 l_c/v_A$). This period of rapid energization suggests that a small number of particles experience some new phenomenon. This number must be small because the energy gained by such particles is visibly larger than the typical deviation from the mean (shadowed area). 

We notice that, superposed on the main regular energy growth, there remain visible smaller-scale oscillations of the particle energy (see inset in Figure \ref{fig:Expgrowth}). In particular, we visually identify a smaller-scale oscillation whose period is $T \simeq 5\times 10^{-2} l_c/v_A$, which corresponds to the particle gyromotion $\tau_g=2\pi/\Omega_g=2\pi r_g/c$. A larger-scale modulation with period $T \simeq 5 \times 10^{-1}l_c/v_A$ is also observed, and this may be correlated with fluctuations of the magnetic field intensity on this timescale. This is suggestive of the simultaneous presence of additional processes, such as mixing of second-/first-order processes and the role of mirroring or drifts. The multiscale complexity of the overall energization dynamics is evidenced by the appearance of at least four timescales in Figure \ref{fig:Expgrowth} ---the exponential time scale, the gyromotion, the modulation seen in the inset, and the second-order energy gain seen to the left and right sides of the exponential phase.

\begin{figure*}[!thb]
\centering
\begin{minipage}{0.48\textwidth}
\includegraphics[width=\columnwidth]{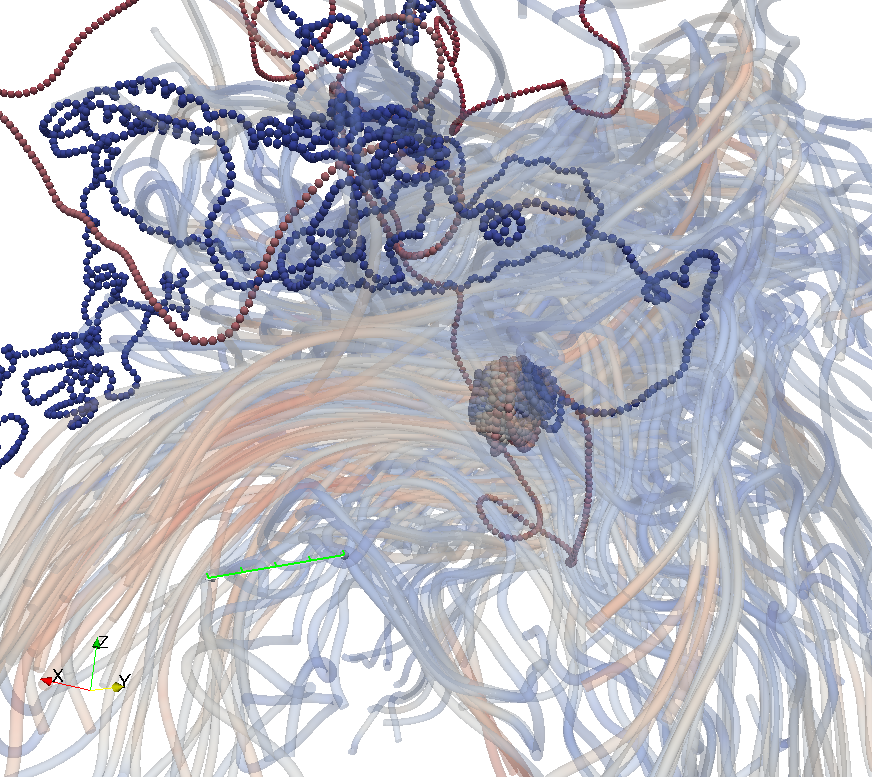}
\end{minipage}
\begin{minipage}{0.48\textwidth}
\includegraphics[width=0.96\columnwidth]{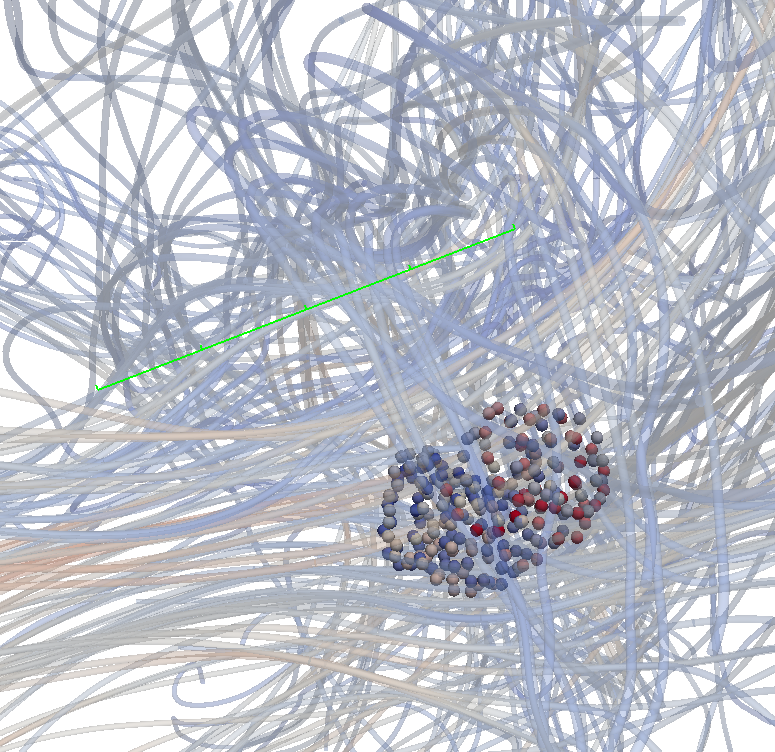}
\end{minipage}
\caption{Particle trajectory in the 3D domain, with the points colored with the particle energy, where the color scale goes from blue to red as the particle energy increases. Magnetic field lines, colored with the magnitude of the magnetic field itself (again from blue to red as the magnetic field magnitude increases), indicate that the particle is trapped in a flux tube and that it is accelerated when the flux tube is feeling the gradients associated with the interaction with another large-scale structure. The right panel shows an inset of the left plot zoomed in the trapping region and limited in time to a few particle gyrations. The green line in each panel corresponds to the correlation length $l_c$.}
\label{fig:3Dplots}
\end{figure*}

The peculiar behavior of the particles during the exponential phase is best illustrated in the bottom panel of Figure \ref{fig:Expgrowth}, where we show the particle's trajectory. One can see that, during the stage of exponential energy growth, the particle is trapped in a small region of the computational domain of size $\sim 0.5 l_c$. In fact, the spatial excursion per unit $l_c/v_A$ is about 10 times smaller between the dashed lines than outside that region. The particle escapes from the trapping region when its gyroradius becomes comparable with the island size $l_{\rm isl}\sim l_c$. 

The phenomenology of this trapping can be also appreciated by looking at the particle trajectory in the 3D domain. Figure \ref{fig:3Dplots} shows the particle trajectory as dots colored with the particle energy, where the color scale goes from blue to red as particle energy increases. Magnetic field lines near the trapping region are also displayed, colored with the amplitude of the field itself (again going from blue to red as the magnetic field amplitude increases). When the particle is not trapped, it carries out an erratic motion in the whole computational domain, akin to an unconstrained random walk. The trapping is associated with a spherical-like motion constrained within a flux-tube-like structure. The energization occurs when the flux tube is perturbed by another large-scale structure, more easily appreciated in the right panel of Figure \ref{fig:3Dplots}. This confirms the scenario that an intense acceleration can occur when magnetic islands and, more in general, large-scale plasma structures, are interacting (collapsing, merging, etc) with other similar structures \citep{DrakeEA06,KowalEA11}, leading to a locally strong magnetic field gradient. 

It is important to point out here that there is no evident association of the structure responsible for the exponential growth of the particle energy with the process of magnetic reconnection. Magnetic reconnection is a sufficient condition for generating large-scale islands where particles can be trapped. Indeed, it can be expected that when the magnetic field reconnects in a turbulent environment, the magnetic islands produced by reconnection interact, thus allowing an intense and fast energization process. However, it is apparently not necessary that reconnection be present during the energization process itself. Other configurations without the explicit invocation of magnetic reconnection, such as the interaction of two large-scale turbulent structures (e.g. flux ropes, as recently reported in recent Parker Solar Probe observations by \citet{pecora2021parker}), may provide a similar behavior, provided that the magnetic geometry of the interaction region favors particle trapping. We remark that the direct acceleration due to the electric field at the reconnection site is negligible for the relativistic particles considered in the present work, given that such particles have a gyroradius much larger than the typical width of current sheets.

\subsection{Characterization of trapping and concomitant energy gain}
\label{sec:character}

In order to characterize the coherent structure that entraps and gives a significant boost to the particle energy, we calculate the current density ${\bm j}=\nabla \times {\bm B}$ and the normalized magnetic helicity $h_m={\bm a}\cdot{\bm B}/(|{\bm a}||{\bm B}|)$, with ${\bm B}=\nabla \times {\bm a}$, interpolated at the particle position. The current density is a direct proxy of the small-scale gradients of the magnetic field, and an intense current density is expected to highlight small-scale structures and current sheets where magnetic reconnection and, in general, dissipative processes may occur \citep[see][and references therein]{pezzi2021dissipation}. On the other hand, magnetic helicity measures the topology of the magnetic field: In particular, a nonnull magnetic helicity indicates twisted, helical magnetic structures.

\begin{figure}[htp!]
  \centering
  \includegraphics[width=\columnwidth]{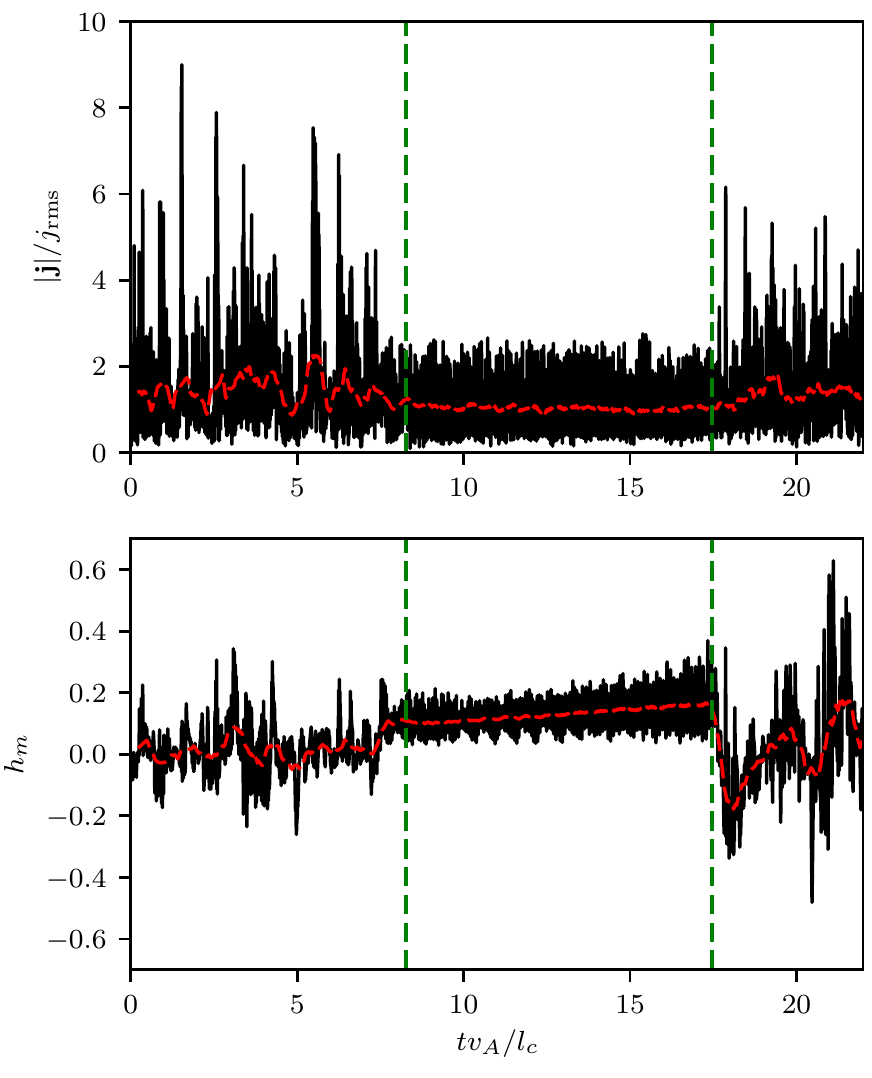}
  \caption{Current density ${\bm j}=\nabla \times {\bm B}$ (top), scaled to its rms value, and normalized magnetic helicity $h_m={\bm a}\cdot{\bm B}/(|{\bm a}||{\bm B}|)$ (bottom), computed at the particle position, as a function of time. The red dashed line corresponds to a large-scale current average performed over $\Delta t\simeq0.5 l_c/v_A$. }
  \label{fig:Structure}
\end{figure}

The values of these variables at the particle position are displayed in Figure \ref{fig:Structure}. The exponential phase is limited by the green dashed vertical lines. The structure responsible for the exponential growth of the particle energy is a relatively quiet region in which the current density is relatively smooth. In comparison, the current outside the structure 
easily reaches intense values $j\gtrsim 4 j_{\rm rms}$, but such intense peaks are not evident within the structure. The magnetic fluctuations are also less intense within the structure, as the rms value of magnetic fluctuations is reduced there by a factor of $3-4$ with respect to the global value.

The structure is furthermore characterized by a finite magnetic helicity, suggesting a flux-tube and/or plasmoid-like shape where magnetic field lines wrap helically on themselves. A finite magnetic helicity also suggests that the structure tends to be force free as ${\bm a}|| {\bm b} \rightarrow {\bm j}|| {\bm b}$, i.e. it may be a large-scale quasi-equilibrium structure typical of intermittent plasma turbulence, where nonlinearities are depleted \citep{MatthaeusEA15}. 

\begin{figure}[htb!]
  \centering
  \includegraphics[width=\columnwidth]{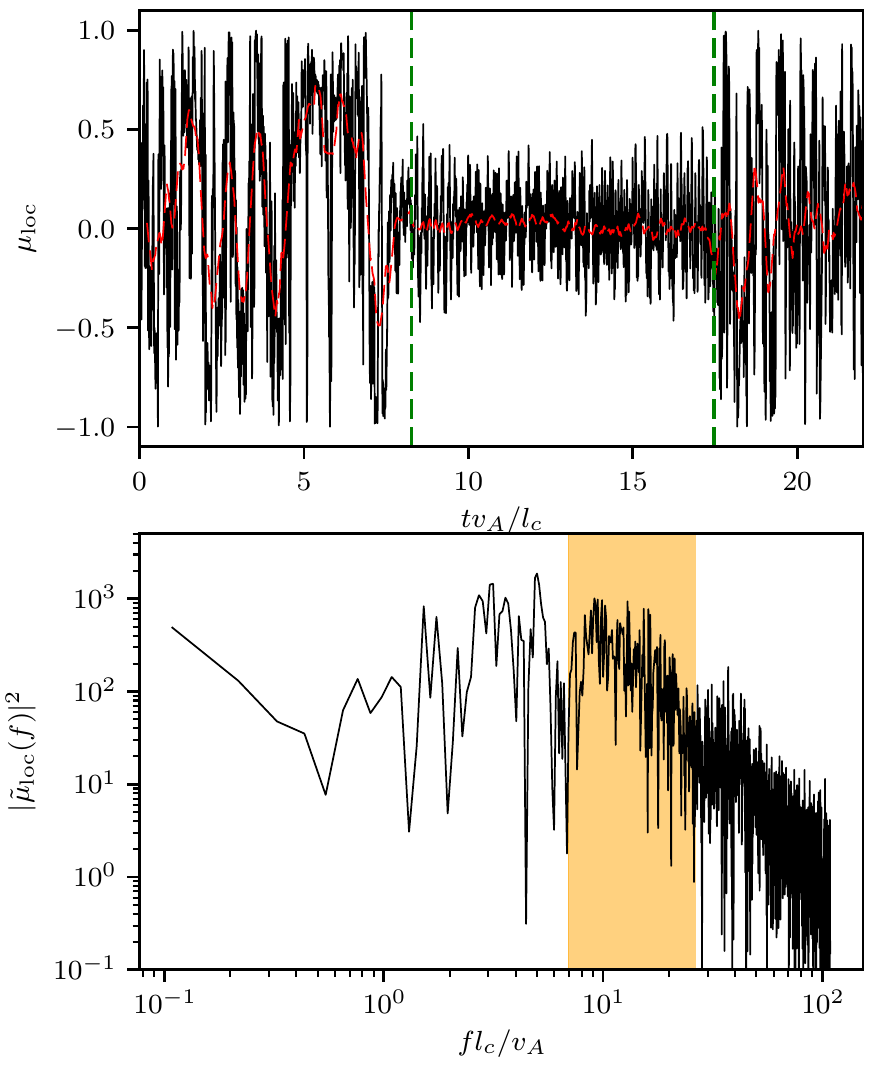}
  \caption{Top: time evolution of the pitch-angle cosine $\mu_{\rm loc}$. The time window associated with trapping is limited by the vertical green dashed lines. The red dashed line is the large-scale current average performed over $\Delta t\simeq0.5 l_c/v_A$. Bottom: time spectrum of the $\mu_{\rm loc}$, when the particle is trapped. The yellow shaded area corresponds to the gyroradius range experienced by the particle during this time window. }
  \label{fig:muloc}
\end{figure}

The properties of the particle trapped in the accelerating coherent structure are also remarkable. The top panel of Figure \ref{fig:muloc} illustrates the pitch-angle cosine of the particle, here defined as
\begin{equation}
  \mu_{\rm loc} = \frac{{\bm B}\cdot {\bm v}}{|{\bm B}||{\bm v}|} = \cos \theta_{{\bm v}{\bm B}},
\end{equation}
because the regular field is absent. 

Concurrently with the period when the particle is trapped, its pitch-angle cosine displays reduced oscillations around the mean zero value. In fact, $\mu_{\rm loc}$ oscillations, estimated as $(\Delta \mu_{\rm loc})_{\rm rms}$, are weaker by a factor of $3-4$ inside the structure with respect to outside. 

The trapped particle has a peculiar motion configuration, with a pitch angle almost perpendicular to the local field. This indicates that the particle is trapped within an elongated 2D-like flux tube and, in particular, it moves in the plane perpendicular to the tube axis. As a consequence, within this period, the particle mainly experiences mainly a perpendicular energization. The evolution of the particle magnetic moment (not shown here) also supports this view because it shows a secular growth on the same timescale as the exponential energy growth of the particle, while the magnetic field is roughly constant within the same time window. Similar observations have been pointed out in the different contexts of the so-called ``second stage'' of acceleration of nonrelativistic particles \citep{DalenaEA14} and, more recently, of electron acceleration in nonrelativistic plasma turbulence \citep{trotta2020fast} and particle acceleration in relativistic plasma turbulence including radiative losses \citep{comisso2021pitch}.

To get insights into the nature of $\mu_{\rm loc}$ oscillations in the time period corresponding to trapping, the bottom panel of Figure \ref{fig:muloc} displays the Fourier time spectrum of $\mu_{\rm loc}$, performed in such a window. The dashed yellow area corresponds to the frequencies associated with the particle gyroradius, which changes within the period due to the particle energization. High-frequency $\mu_{\rm loc}$ oscillations, associated with the particle gyromotion, are combined with lower-frequency fluctuations, possibly related to smaller-scale turbulent fluctuations or other effects, such as mirroring and drifts.

\begin{figure}[htb!]
  \centering
  \includegraphics[width=\columnwidth]{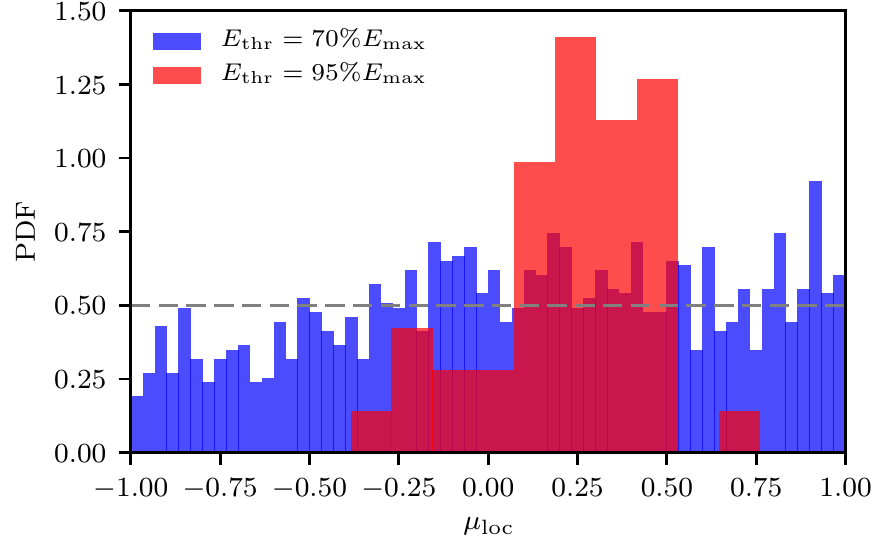}
  \caption{PDFs of $\mu_{\rm loc}$ conditioned with the particle energy for the top $30\%$ of the most energetic particles (blue) and for the top $5\%$ (red), with the maximum energy being $E_{\rm max}=24.4\, {\rm PeV}$ ($r_{g,{\rm max}}/l_c=2.2$). The horizontal dashed gray line displays the $1/2$ value, corresponding to isotropic distribution. The exponential acceleration responsible for the high-energy tail is associated with a small pitch-angle cosine $\mu$. Because we are not removing particles once they enter the exponential phase, they consequently undergo the standard pitch-angle process, thus producing the observed spreading in the $\mu_{\rm loc}$ distribution.
}
  \label{fig:PDFmu-Econd}
\end{figure}

To demonstrate that the most intense energization is statistically associated with a small pitch-angle cosine, Figure \ref{fig:PDFmu-Econd} displays the PDFs of the pitch-angle cosine conditioned to the particle energy. In particular, we computed the PDFs by considering the full ensemble of particles up to the time $T\simeq 22 l_c/v_A$ and by setting the threshold $E_{\rm thr}=70\% E_{\rm max}$ (blue) and $E_{\rm thr}=95\% E_{\rm max}$ (red), where 
$E_{\rm max}=24.4\, {\rm PeV}$ (i.e. $r_{g,{\rm max}}/l_c=2.2$). Particles below the threshold show a distribution compatible with isotropy, with a mean pitch-angle value of $1/2$, reported in Figure \ref{fig:PDFmu-Econd} with a gray dashed line. On the other hand, the most energetic particles display a strongly anisotropic distribution, peaked at small $\mu_{\rm loc}$, becoming more evident for larger thresholds. 

This confirms that the most-energetic particles preferentially have a small local pitch-angle cosine, i.e. that they move perpendicular to the local magnetic field. As we discuss in \S \ref{sec:discussion}, this may be the very cause of the trapping: The velocity of the particles parallel to the local field is such that the particle stays in the structure long enough to be energized (while possibly drifting) by gradients in the magnetic field. During this process, the particle pitch angle changes gradually, as would be expected due to the quasi-conservation of the adiabatic invariant: Because the field changes along the trajectory (at most by the $\delta B/B_{\rm rms}$ calculated at the scale of the Larmor radius of the particles) and $v_\perp p_\perp/B$ must stay constant, then $p_\perp$ must change, hence the pitch angle changes \cite[]{volk1975} by about $\delta B/B_{\rm rms}$. This leads to a spread in the PDF of the pitch-angle distribution of the most energetic particles.

\section{Discussion of the physical processes at work}
\label{sec:discussion}

Here we discuss our results and their implications for astrophysical systems. It has been found that relativistic particles moving in stationary turbulent electric and magnetic fields, constituted by a snapshot in time of the incompressible MHD simulations, experience a complex energization process owing to the turbulent inductive electric field. The character of the process reveals at least a twofold nature. The second-order, stochastic acceleration affecting the whole ensemble of particles goes hand in hand with a first-order mechanism, impacting only a few particles that are trapped in large-scale coherent structures of turbulence. These trapped particles experience exponential energy growth until the time when the gyroradius becomes comparable with the transverse size of the structure.

The phase of exponential increase in particle energy is of special interest for its potential implications for astrophysical systems. As discussed in previous sections, this stage is rather complex: It requires that a particle enter a region where two large-scale structures seem to be interacting, thereby leading to gradients in the induced electric field and most likely in the magnetic field as well. The number of particles showing evidence of this exponential energy increase is very small, only a few out of $10^5$ ($0.001\%$). All these particles seem to have a very small pitch-angle cosine, namely $\mu\sim 0$. They also exhibit a similar phenomenology in terms of characteristic times: The trapping time is quite large ($\sim 5-10 l_c/v_A$), and various oscillations, as described in the previous section, are recovered. The formation of such structure in a generic turbulence box is difficult to characterize in a quantitative way, also due to intermittency: The standard Kolmogorov scalings could be locally violated due to spatial inhomogeneities \citep{MatthaeusEA15}. In particular, magnetic fluctuations could become weaker in large-scale coherent structures with respect to the globally averaged values because nonlinearities could be partially suppressed within large-scale structures, and perhaps elevated near flux-tube boundaries. In fact, as discussed earlier in this article, the structure in which the exponential acceleration takes place does not show evidence for special activity: it might have originated from a reconnection event, but it is not a site of reconnection. Indeed, reconnection plays no role in causing the rapid energy increase that we observe in the simulations. 

Figure \ref{fig:3Dplots} illustrates very clearly how complex the region is where the particle energy is seen to increase exponentially. The only properties that we can confidently associate to this region are (1) the presence of a rather organized large-scale flux tube that seems to be interacting with another structure, and (2) a gradient in the local electric field, due to plasma motion. Moreover, it is reasonable to speculate that there may be a gradient in the magnetic field due to the interaction between structures. 

Given the complexity of the situation, it may be helpful to use a toy model to illustrate the different effects and check whether the qualitative picture is reproduced. The toy model gives us the opportunity to comment on the different physical processes at work. 

\begin{figure}[htb!]
  \centering
  \includegraphics[width=\columnwidth]{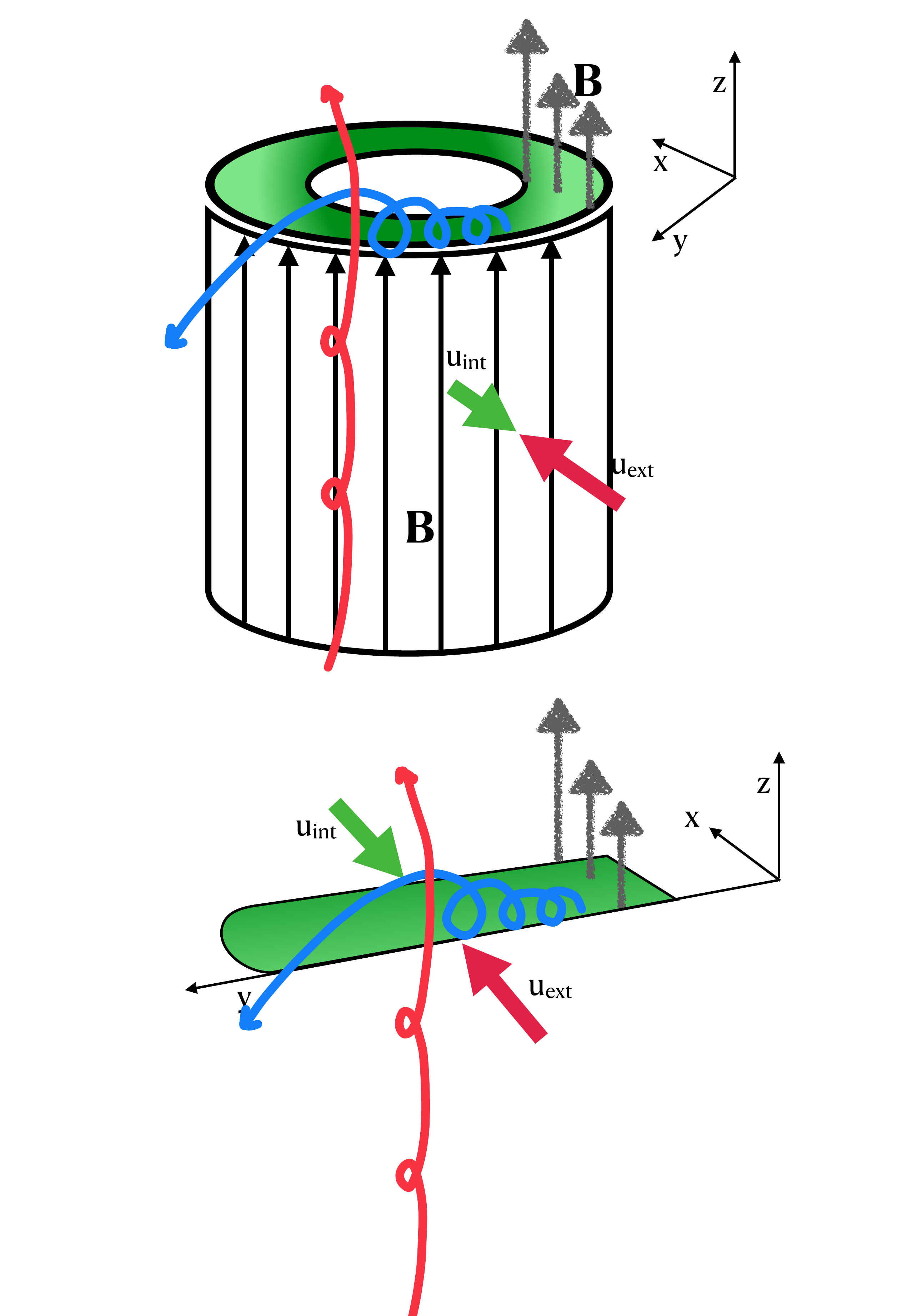}
  \caption{Sketch of the geometry of the flux tube interacting with another structure (top part). At the bottom is shown a zoom-in into the region where the gradient of plasma velocity is present.}
  \label{fig:toy}
\end{figure}

We show the geometry of our toy model in Figure \ref{fig:toy}: The magnetic field ${\bm B}$ in that region is assumed to be oriented in the $\hat z$ direction, but it is assumed to have a gradient in the direction of the center of the tube, within a ring (green region) of size $L_{\rm grad}$. We also assume that at least on a fraction of the surface of the tube, the plasma velocity has a gradient along the $\hat x$ direction (see zoom-in in the lower part of Figure \ref{fig:toy}). In the figure, this gradient is shown in the form of a sign reversal of the velocity but this is not required. We will comment below on the implications of a fluid velocity crossing the value $u=0$.

Let us start by discussing the role of a gradient in the magnetic field, although the existence of a strong gradient does not emerge in an evident way from the simulations: A magnetic field, with or without gradients, cannot make work on charged particles. Hence, the exponential energy increase is certainly not related to such a gradient. On the other hand, the gradient introduces a drift, whose direction depends on the direction of the gradient. With reference to the situation illustrated in the lower part of Figure \ref{fig:toy}, this grad-B drift leads the particle to move in the $y$ direction, namely to stay in the green-colored region. The drift velocity is proportional to the gradient: 
\begin{equation}
    {\bm v}_D\simeq\frac{pvc}{2q}\frac{{\bm B} \times {\nabla}B}{B^3}\approx 
    \frac{1}{2}\frac{r_L(p)v}{L_{\rm grad}} \hat y,
    \label{eq:vD}
\end{equation}
namely, the lower the gradient, the lower the drift velocity, so that the particle can stay longer in the island. Note that Equation (\ref{eq:vD}) includes the grad-B drift, while the curvature drift is subdominant because we showed that the motion occurs at $\mu \sim 0$, i.e. $p_{\rm perp}\simeq p$. 

Inspired by the results of our simulations, we first assume that the particle has a very small pitch-angle cosine $\mu$, namely, its motion is constrained to be in the $x-y$ plane (blue curve in Figure \ref{fig:toy}). A small value of $\mu$ means that the particle can travel in the $\hat z$ direction with velocity $\sim c\mu$ (all of the particles in the simulation are relativistic) and will eventually leave the island in a time $\sim L_{\rm isl}/c\mu$. For $\mu\sim 1$ (red curve in Figure \ref{fig:toy}), the escape time would be exceedingly fast and no appreciable acceleration can take place (see below). 

At the same time that the particle rotates in the $x-y$ plane, it is advected with the local plasma, which is expected to move at speeds close to the local Alfv\'en speed $v_A$. The region of the gradient in the plasma speed (which corresponds to the gradient in the induced electric field) can then be crossed in a time $L_{\rm grad}/v_A\sim \eta L_{\rm isl}/v_A$, where we assumed that the gradient develops on a fraction $\eta$ of the size of the island, $L_{\rm grad}\sim \eta L_{\rm isl}$. The time to escape the island along $\hat z$ exceeds the time scale of advection if $\mu\eta \ll \frac{v_A}{c}$. This apparently simple and rather constraining conclusion on the pitch angle of the particles in the acceleration region is in fact affected by several phenomena that can possibly enhance the trapping: one such phenomenon is diffusion, but it is hard to imagine that it may be effective. In fact, the particles that we are considering have an initial Larmor radius of only one order of magnitude smaller than the coherence scale, which in turn is of the same of order as the size of the island, $L_{\rm isl}$. Hence, the escape time is a fraction of $l_{c}^2/(c l_c/3)\sim 3l_c/c$, where we assumed that the diffusion coefficient is not too far from that expected at $r_L\sim l_c$ (see Figure \ref{fig:Diso-sat}). The time estimated in this way is very close to the ballistic time scale (within an order of magnitude), too short to be of any relevance for CR trapping. Moreover, the fact that the pitch angle is close to zero makes diffusion ineffective, because the resonance condition would require modes with a very large wavenumber $k$, which is not accessible in our simulations, and not abundant in the standard turbulence spectrum anyway, especially after accounting for anisotropy in the cascade. 

A second phenomenon that is instead expected to be more effective is associated with the existence of perturbations in the magnetic field along the $\hat z$ direction: Because on the scale of a few gyrations it is a good approximation to assume that the quantity $v_\perp p_\perp/B$ is conserved, if there are fluctuations of order $\Delta=\delta B(1/r_L)/\delta B(1/L_{\rm isl})$ in the local field, the conservation of this quantity also implies that $p_\perp$ must be changing, namely $\mu$ must be changing. Oscillations in $B$ imply oscillations in the pitch angle of magnitude $\propto \Delta^{1/2}$ \cite[]{volk1975}. Unfortunately, for the parameters of our simulation, the quantities $\Delta^{1/2}$ and $v_A/c$ are too close to identify this effect unambiguously. However, for more realistic values of the ratio $v_A/c$, the effect of oscillations associated with longitudinal gradients in the magnetic field should dominate and force particle trapping provided that $\mu$ is smaller than $\Delta$. 

We finally come to the effect of the gradient in the plasma velocity. If there were a homogeneous plasma velocity $u$ in the $\hat x$ direction, this would result in an induced electric field $E_y=\frac{u(x)}{c}B$, directed in the $\hat y$ direction. Notice that as long as the plasma speed is spatially constant, one can always move into a reference frame in which this velocity is absent: Hence, the presence of this induced electric field only introduces a drift velocity in the $\hat x$ direction and no net increase in the energy of the particles can exist. In other words this drift is only the ${\bm  E}\times {\bm B}$ drift and leads to the particle advection with the background plasma in the $\hat x$ direction. The case where the velocity of the plasma is not constant in the $x$ direction is more interesting. Notice that the simulations used here are incompressible, hence $\nabla\cdot {\bm u}=0$. This does not contradict the assumption that a gradient $du/dx\neq 0$ exists, for instance, if two structures are moving against each other at $u\sim v_A$. 

Below we show that the existence of this drift is the very source of particle acceleration. We will reach this conclusion in two independent ways.  

The evolution in time of the particle distribution function $f$ of the particles, under the effect of the $du/dx$ gradient alone, can be written as:
\begin{equation}
\frac{\partial f}{\partial t}-\frac{1}{3}\frac{du}{dx}\frac{\partial f}{\partial \xi}=0,
\label{eq:transport}
\end{equation}
where $\xi=\ln(p)$, and we used $du/dx\approx v_A/L_{\rm grad}$. If the two structures are moving against each other, a more likely estimate for the gradient would be $2v_A/L_{\rm grad}$, but the difference is only quantitative, not qualitative. Using the method of characteristics, one gets
\begin{equation}
    \frac{dp}{p}=\frac{1}{3}\frac{du}{dx}dt \to p(t)\propto \exp\left[ \frac{v_A t}{3 L_{\rm grad}}\right].
\label{eq:exp1}
\end{equation}
The particle momentum is expected to grow exponentially due to the presence of a difference in velocity felt by particles during gyration around the magnetic field. One can look at this phenomenon in at least two independent ways: One way is to imagine that in each particle orbit the electric field in one half is not exactly compensated by the second half, hence there is a net electric field that can energize the particle. The second way to look at this as a first-order Fermi process, in which the particle bounces (each half-orbit) on a fluid moving with a different speed, qualitatively similar to what happens close to a shock front, where however the gradient is much larger. The acceleration has to come to an end when the gyroradius becomes comparable with the size of the trapping region unless more constraining phenomena occur on shorter time scales. 

Notice that the factor of $1/3$ in Equation (\ref{eq:transport}) is derived under the assumption of particle isotropy, which clearly does not apply here. We expect that the time scale of acceleration of particles with $\mu\sim 0$ must be somewhat shorter than the $3L_{\rm grad}/v_A$ suggested by Equation (\ref{eq:exp1}).

In order to prove that our conclusion is solid, we also derive an analogous result in a more formal way, starting from the equation of motion of a particle moving in a setup as in Figure \ref{fig:toy}. In this derivation, we ignore the effect of particle drift due to the gradient of the magnetic field, hence
\begin{eqnarray}
\frac{d p_x}{dt}=\frac{q}{c} v_y B \label{eq:motionx} \\
\frac{d p_y}{dt}=-\frac{q}{c} v_x B +q \frac{u(x)}{c} B,
\label{eq:motiony}
\end{eqnarray}
where we used the expression for the local induced electric field derived above. Recalling now that ${\bm p} = m \gamma {\bm v}$, where $\gamma$ is the Lorentz factor of the particle, and multiplying Equation (\ref{eq:motionx}) by $v_x$ and Equation (\ref{eq:motiony}) by $v_y$, we can deduce that
\begin{eqnarray}
\frac{1}{2}m \gamma \frac{d v_x^2}{dt}+ m v_x^2 \frac{d\gamma}{dt}=\frac{q}{c} v_x v_y B \\
\frac{1}{2}m \gamma \frac{d v_y^2}{dt}+ m v_y^2 \frac{d\gamma}{dt}=- \frac{q}{c} v_x v_y B +\frac{q}{c}u(x)v_y B.
\label{eq:motion1}
\end{eqnarray}
Summing all terms and recalling some basic relations of relativistic kinematics, one obtains
\begin{equation}
    \frac{dp}{dt}=\frac{q}{c} B v_y u(x).
    \label{eq:motion2}
\end{equation}
At this point we assume that the particle velocity is already relativistic, $v\simeq c$, and that the particle trajectory is weakly modified by the gradient in $u(x)$. This means that the gradient is assumed to be weak on the scale of the Larmor gyration of the particles. In this case, Equation \ref{eq:motion2} can be averaged over many gyrations and recalling that $x(t)\propto sin(\Omega t)$ and $v_y(t)\propto sin(\Omega t)$, the average leads to 
\begin{equation}
    \left\langle \frac{dp}{dt}\right\rangle = \frac{1}{2} a p,
\end{equation}
where $a\approx v_A/L_{\rm grad}$ is the gradient of $u(x)$. It follows that 
\begin{equation}
    \langle \frac{dp}{p}\rangle = \frac{1}{2} \frac{v_A}{L_{\rm grad}} dt \to p(t)\propto \exp\left[ \frac{1}{2} \frac{v_A}{L_{\rm grad}} t \right].
\end{equation}
This result is formally the same as in Equation (\ref{eq:exp1}), but it correctly shows that the time for exponential increase of the momentum is slightly shorter than expected for an isotropic particle distribution. 
In the absence of escape mechanisms, the trapping is expected to cease when the particle gyroradius becomes comparable with the island size, $r_L\sim L_{\rm isl}$. This provides a characteristic time for the duration of the energization process,
\begin{equation}
t \approx \frac{2 L_{\rm grad}}{v_A} \Lambda,
\label{eq:Texp}
\end{equation}
where 
$r_{L,0}=p_0 c/e B$ is the initial particle Larmor radius and $\Lambda=\ln{\left(\frac{L_{\rm isl}}{r_{L,0}}\right)}$.

In order to have effective acceleration, the acceleration time must be shorter than all the time scales of escape parallel to $B$ and due to drifts. 
By imposing that the time of escape along $\hat z$, $\tau_{{\rm cross},||}=L_{\rm isl}/c\mu$, be longer than the acceleration time, we obtain the constraint
\begin{equation}
    \mu \lesssim \frac{v_A}{c} \frac{L_{\rm isl}}{2 L_{\rm grad}} \frac{1}{\Lambda}.
\label{eq:mulim}
\end{equation}
This result provides us with a simple explanation of the reason why only particles with a small value of $\mu$ actually show evidence of exponential momentum increase.

The issue of drifts is more subtle: the ${\bm E}\times {\bm B}$ drift, as discussed above, simply leads to particle advection along the ${\bm x}$ direction. The region of the gradient is then crossed in a time of order $\tau_{{\rm esc},x}\sim L_{\rm grad}/v_A$, comparable to, though shorter than, the acceleration time. However, if the two islands move against each other with roughly the same speed, $\sim v_A$, the guiding center of the particle may be advected at speeds much smaller than $v_A$, as a result of the fact that $u(x)$ crosses zero. 

The drift along the $\hat y$ direction is due to the gradient in the magnetic field and in principle can be very fast because the drift velocity is given by Equation (\ref{eq:vD}), so that the time to drift out of the region where the gradient exists is of order 
\begin{equation}
    \tau_{{\rm esc},y}\simeq \frac{L_{\rm grad}^2}{\frac{1}{3}r_L v}\simeq 
    \frac{3\eta^2L_{\rm isl}}{0.1 c},
\end{equation}
where we assumed that $r_L\sim 0.1 L_{\rm isl}$. For the parameters adopted in the simulation, this time is $\tau_{{\rm esc},y}\sim 0.3 L_{\rm isl}/c$, much shorter than all other times scales. In the presence of such an effect, no appreciable acceleration should be expected. The evidence of an exponential increase in particle energy suggests that either the gradient in the magnetic field is a small fraction of $B/L_{\rm grad}$, or that it is present all along the surface of the flux tube (green region in the top part of Figure \ref{fig:toy}), so that the particle drifts around the tube while retaining its pitch angle, as discussed above. 

It would be very useful to further investigate the phenomena discussed above, but because in our simulations $v_A=c/20$, such time scales  are too close to each other to reach a definitive conclusion on the hierarchy of drift timescales. 

We conclude our in-depth discussion of the physical processes at work in the acceleration region by commenting on the role of reconnection. As we discussed in \S \ref{sec:character}, our simulations suggest that the region where particle energization is fast is not very active: It may be the result of a reconnection event, but it is not a reconnection region. This is also to be expected: The current sheet is very thin compared with the Larmor radius of the relativistic particles considered here, hence the resistive electric fields cannot be responsible for particle acceleration. In this sense, a model like the one of \cite{degouveiadalpino2005production}, in which the escape from the region is regulated by the speed of the exhaust of the reconnection phenomenon, appears to be not well justified. On the other hand, \cite{KowalEA11} correctly pointed out that interacting islands away from reconnection events can energize particles, even exponentially. This result was, however, obtained in simulations that were optimized to create reconnection regions. We show that even in a generic 3D MHD simulation box there are structures that lead to the same physical phenomenon, and we provide a physical explanation of the phenomenon and a recipe of the conditions required for the acceleration to take place. It is also worth noticing that a first-order particle acceleration in converging islands in relativistic turbulence was also found by \cite{ComissoSironi18}. The case of relativistic particles moving in a relativistic plasma $v_A\sim c$ may be considerably different from the one described above.

The final part of this section is devoted to a discussion of the possible relevance of these phenomena for astrophysical turbulence, for instance, in the Galaxy as a whole. 

In principle the exponential increase associated with phenomenon of particle trapping in interacting islands may be of great importance: For instance, if we take the turbulence in the Galaxy as our laboratory, one would expect $v_A\sim 10\, {\rm km}\, {\rm s}^{-1}$ and $L_{\rm isl}\sim\,$few tens of parsec, with a typical magnetic field of $3\,\mu$G. If a CR particle enters one such structure and suffers an exponential increase in energy up to the point where the Larmor radius equals $L_{\rm isl}$, a maximum energy of $\sim 20$ PeV would be reached, tantalizingly close to the energy of the knee. This simple numerical estimate stimulates some additional questions: What is the time scale for such acceleration? And what is the probability that a CR particle may encounter such a peculiar structure before escaping the Galaxy? 

The time scale for exponential increase is as in Equation (\ref{eq:Texp}): One can see that for realistic values of the parameters, the acceleration time is $\sim (1-10) l_c/v_A$, where we assumed that $L_{\rm isl}\sim l_c$. This seems to be in accordance with the numerical results shown in Figure \ref{fig:Expgrowth} (top panel). The fact that this time is comparable to or exceeds the eddy turnaround time $l_c/v_A$ is a source of concern because both the simulations and the toy model discussed above assume that the turbulence is static (the propagation of test particles was carried out in a snapshot at a given time of the MHD simulation). 

For turbulence in the Galaxy, the acceleration time estimated above is of order $\sim 1$ million years. For particles at the knee, the escape time from the Galaxy can be deduced from an extrapolation to high energies of the low-energy ($\lesssim 1$ TeV) confinement time as inferred from secondary/primary ratios \cite[]{Evoli2019} and from the $Be/B$ ratio, as discussed recently by \cite{Evoli2020}, using AMS-02 data. A reasonable estimate for such escape time at $E\sim\, {\rm PeV}$ is of $\sim 0.5$ Myr, comparable with the acceleration time. The situation can be considered somewhat more promising if one thinks that lower-energy particles (with longer confinement time) are the ones required to be trapped in the interacting islands and eventually getting energized. 

Assuming that in a time of the order of the confinement time a CR particle can probe the statistical properties of the turbulence, the question arises of how many particles can potentially interact with a region where trapping occurs and particle energy increases exponentially. In the simulations we ran, only a few particles out of the $10^5$ ($0.001\%$) experienced exponential energy increase. The probability of order $10^{-5}\div 10^{-4}$ that one of the particles, while carrying out a diffusive motion, may encounter a region of size $\sim l_c$ in which there are the right conditions for trapping to occur is hard to quantify in that it is the convolution of turbulence properties (volume filling factor of islands that interact in the right way, in the presence of intermittency) and properties of the particle trajectory at the time of entering the island: As we discussed above, only particles that approach the region with very small $\mu$ can be trapped in the region for long enough to experience the exponential energy increase. 

Because diffusion isotropizes the particle distribution function (outside the island), the fraction of particles that at any point in the box have a pitch-angle cosine $\mu\sim v_A/c$ is $\sim \frac{1}{2}v_A/c = 0.025$, where the numerical value refers to the conditions of the simulation, while in the Galaxy that number would be $\sim 1.7\times 10^{-5}$. The fact that only a few particles out of $10^5$ suffer exponential energy increase implies that, in the simulation, the filling factor of the island that allows such phenomenon is very small, of the order of $\sim 10^{-3}$. 

A dedicated investigation of these issues would be most important as a future development of the concepts discussed in the present article and is crucial to assess the importance of these phenomena for particle acceleration in nature.

\section{Conclusions} \label{sec:conclusions}
The use of high-resolution magnetohydrodynamic simulation data in conjunction with orbit calculations of a large number of charged test particles enables detailed examination of both 
spatial transport and energization in a generic 3D incompressible turbulence simulation not specifically devised to study reconnection. 
The limited number of scales that can be simulated in 3D at this time makes it difficult to go much beyond the state of the art in terms of investigating particle transport: Nevertheless we determined the diffusion coefficient of relativistic particles in a range of scales where resonances are numerically accessible in the simulation. This bounds us to about one decade in energy below the energy for which the Larmor radius equals the correlation length $l_c$ of the turbulence. At such energies, it is difficult to spot the effect of anisotropic cascade with respect to the local magnetic field, typical of MHD, although such anisotropy is certainly visible when studied in terms of statistical indicators \citep{matthaeus2012local}. In this sense, our results on the diffusion coefficient are compatible with those of \cite{Cohet2016} but do not add to it. They are also compatible with the results previously obtained by \citet{SubediEA17} and \citet{dundovic2020novel} using synthetic isotropic turbulence, rather than MHD turbulence. This latter result confirms that the effects of anisotropic cascades are not yet visible on the scales accessible to particles. 

Our results on particle energization are much more interesting: We find that the bulk of the test particles simulated here are subject to a secular, second-order acceleration process, due to the interaction of the particles with the random electric fields induced through plasma motion in the simulation. A few out of the $10^5$ test particles for which we simulate the trajectories happen to suffer very fast acceleration---in fact exponential in time. This process is seen to end when the Larmor radius becomes comparable with the size of the magnetic structure in which the particles reside.  

The second-order stochastic acceleration process has been analyzed by employing standard techniques including computation of the running diffusion coefficient in energy space. We find empirical values for energy diffusion $D_{EE}\simeq 0.01 v_A E_0^2/l_c$, and for the associated characteristic time  $\tau_{\rm diff,E} \sim 10^2 l_c/v_A$. These quantities are in qualitative agreement with naive expectations: The energy of a particle is expected to change in a random way (increase or decrease) due to the interaction with induced electric fields, so that their energy changes as $\left(\frac{\xi v_A}{c}\right)^2 E$, with $\xi\lesssim 1$, in a time that is approximately $\sim l_c/c$ (see Figure \ref{fig:Diso-sat}). It follows that the diffusion coefficient in energy can be estimated as $D_{EE}=\left(\frac{v_A}{c}\right)^2 E \frac{c}{v_A}\frac{v_A}{\xi l_c}$. For the $v_A/c=1/20$ adopted in our calculations $D_{EE}\approx \xi^2 0.05 E^2/l_c$. 

We characterize further the second-order acceleration process by evaluating the time-dependent probability distribution functions, for both particle energy and time increments of particle energy. These additional tests show the second-order nature of the phenomenon. 

As mentioned above, a few of the test particles in the simulation appear to have a quite different behavior, in that in addition to the slow second-order process, they happen to encounter special locations in the box where the energy is seen to increase exponentially, for times exceeding or about 10 Alfv\'en crossing times of the magnetic coherence length.

Closer examination reveals that these particles maintain a pitch angle not far from 90$^\circ$ during this period, while their spatial trajectory is highly confined and differs greatly from its more typical random walk nature. In essence, these particles have become temporarily trapped within particular turbulence structures that are characterized by strong gradients. The structures are not directly associated with regions of activity in the plasma, and in fact, if any, they show less than normal activity, although visual inspection of these regions suggests the existence of extended interaction areas with at least another structure. 

We built a toy model that includes in a simplified way the main drifts and gradients in the local electric fields. The toy model shows that the main reason for rapid particle energization is a sort of first-order Fermi process in which the energy of the particles grows due to a local gradient in the plasma velocity field. Both the temporal evolution of the energization process and the typical time scales for particle acceleration, as well as the maximum energy, are qualitatively reproduced in a correct way.

\section{Acknowledgements}
   
P.B. and O.P. are grateful to G. Kowal, L. Sironi, and S. Servidio for very useful conversations on topics related to this article. O.P. also appreciated friendly yet significant discussions with D. Trotta. We also thank Riddhi Bandyopadhyay for his support in making use of the MHD simulations and the anonymous reviewer whose comments definitely improved our manuscript. Test-particle simulations here presented have been performed on the Newton cluster at the University of Calabria (Italy). W.H.M.'s research is partially supported by NSF-DOE Grant PHY2108834 and by the Parker Solar Probe ISOIS project under subcontract SUB0000165 from Princeton. Last but not least, O.P. thanks Maria Pezzi for her precious helpfulness.



%
\bibliographystyle{plainnat}



 \newcommand{\BIBand} {and} 
  \newcommand{\boldVol}[1] {\textbf{#1}} 
  \providecommand{\SortNoop}[1]{} 
  \providecommand{\sortnoop}[1]{} 
  \newcommand{\stereo} {\emph{{S}{T}{E}{R}{E}{O}}} 
  \newcommand{\au} {{A}{U}\ } 
  \newcommand{\AU} {{A}{U}\ } 
  \newcommand{\MHD} {{M}{H}{D}\ } 
  \newcommand{\mhd} {{M}{H}{D}\ } 
  \newcommand{\RMHD} {{R}{M}{H}{D}\ } 
  \newcommand{\rmhd} {{R}{M}{H}{D}\ } 
  \newcommand{\wkb} {{W}{K}{B}\ } 
  \newcommand{\alfven} {{A}lfv{\'e}n\ } 
  \newcommand{\alfvenic} {{A}lfv{\'e}nic\ } 
  \newcommand{\Alfven} {{A}lfv{\'e}n\ } 
  \newcommand{\Alfvenic} {{A}lfv{\'e}nic\ }

\end{document}